\documentclass{INTERSPEECH2023}

\usepackage{stfloats}
\usepackage{caption}
\usepackage{amsfonts}
\usepackage{multirow}

\interspeechcameraready

\title{Adaptive Contextual Biasing for Transducer Based \\Streaming Speech Recognition
}

\name{Tianyi Xu$^*$, Zhanheng Yang$^*$, Kaixun Huang, Pengcheng Guo, Ao Zhang, Biao Li, Changru Chen, Chao Li, Lei Xie$^\dagger$}
\address{
  Audio, Speech and Language Processing Group (ASLP@NPU), School of Computer Science, \\ Northwestern Polytechnical University, Xi'an, China
\email{xutianyi@mail.nwpu.edu.cn, lxie@nwpu.edu.cn}
\thanks{$^*$Equal contribution. $^\dagger$Lei Xie is the corresponding author.}}
\usepackage{threeparttable}
\begin{document}

\maketitle
\begin{abstract}
By incorporating additional contextual information, deep biasing methods have emerged as a promising solution for speech recognition of personalized words.
However, for real-world voice assistants, always biasing on such personalized words with high prediction scores can significantly degrade the performance of recognizing common words.
To address this issue, we propose an adaptive contextual biasing method based on Context-Aware Transformer Transducer (CATT) that utilizes the biased encoder and predictor embeddings to perform streaming prediction of  contextual phrase occurrences. Such prediction is then used to dynamically switch the bias list on and off, enabling the model to adapt to both personalized and common scenarios.
Experiments on Librispeech and internal voice assistant datasets show that our approach can achieve up to 6.7\% and 20.7\% relative reduction in WER and CER  compared to the baseline respectively, mitigating up to 96.7\% and 84.9\% of the relative WER and CER increase for common cases. Furthermore, our approach has a minimal performance impact in personalized scenarios while maintaining a streaming inference pipeline with negligible RTF increase.
  
\end{abstract}

\section{Introduction}

\begin{figure*}[t]
\centering
\resizebox{1.0\textwidth}{!}{\includegraphics{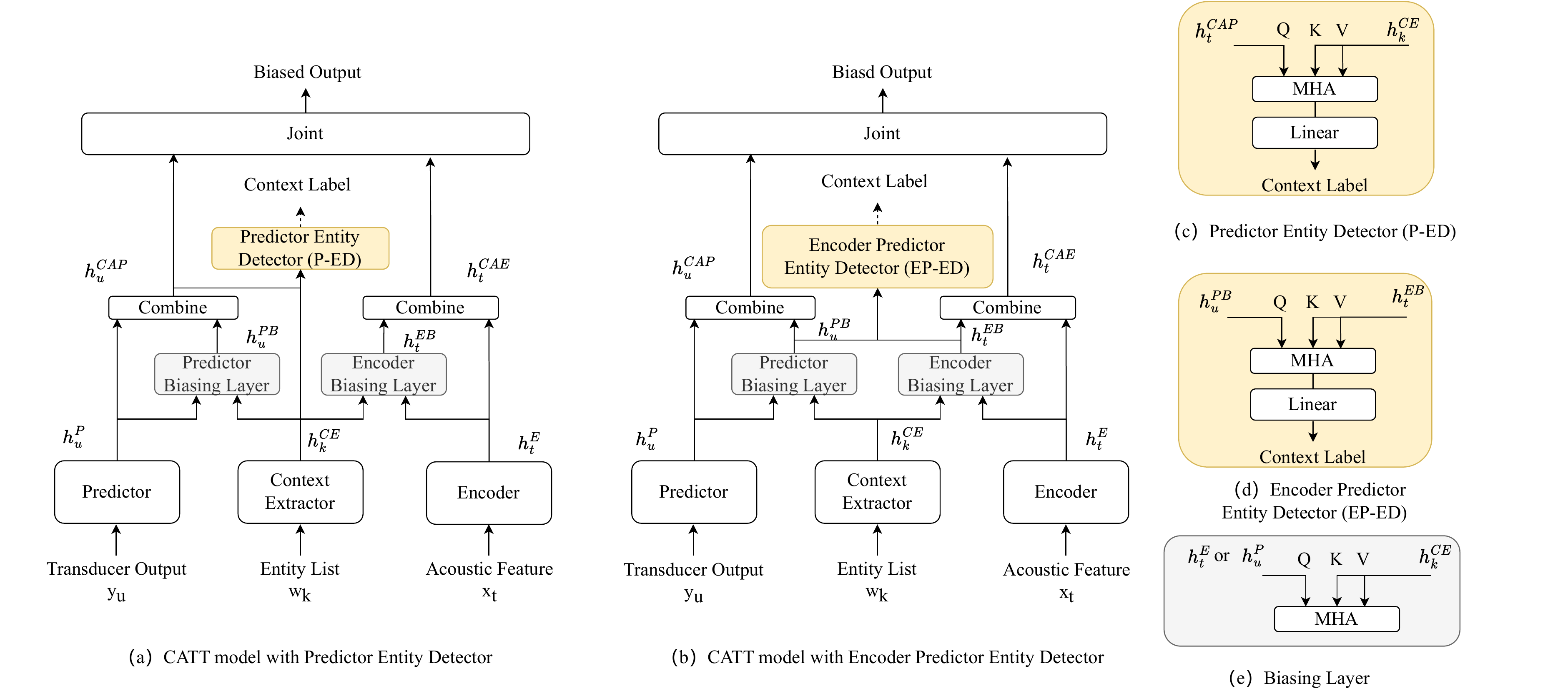}}
\caption{(a) CATT model with P-ED. (b)   CATT model with EP-ED (c) P-ED module  (d)  EP-ED module (e)  Biasing Layer }
\vspace{-0.5cm}
\end{figure*}
Recent advancements in deep learning have propelled end-to-end automatic speech recognition (E2E ASR) systems to new heights, making them the backbone of many speech recognition applications. E2E ASR methods such as connectionist temporal classification (CTC)~\cite{2006ctc, 2014ctc,li2019towards}, recurrent neural network Transducer (RNN-T)~\cite{2019rnnt, 2021rnnt, wang2021efficient}, and attention-based encoder-decoder (AED)~\cite{2015aed, 2014aed, 2016aed}, have been widely adopted. However, since data coverage greatly impacts deep learning techniques, for sentences containing words that rarely occur in training data, ASR performance is often severely impaired. These rare words include but not limited to contact names, location names, internet buzzwords, or product names.

Generally, there are two types of methods that integrate context information into E2E ASR systems. The first method is to compile the biasing phrases into a finite state transducer (FST) and incorporate it into the decoding process~\cite{mohri2002weighted}.
However, the compiling process requires calibration to find the optimal fusion weight, so the adjustment of the context list can be non-trivial. 

The second method is to use encoders to directly inject context information~\cite{2019nnbias, 2021nnbias, 2021nnbias2, 2021nnbias3, 2022nnbias}. Previous works such as CLAS~\cite{2018nnbias} and CATT~\cite{2021nnbias4} use an additional location-aware attention mechanism to leverage contextual information into E2E ASR. 
The introduced entity encoder enables the entity list to be personalized for individual users. However, this personalization comes at a cost: the model has less prior knowledge of the customized words, which can result in false alarms. In other words, the model may mistakenly identify non-entity names as entity terms, leading to a decrease in overall recognition performance, particularly for words that are phonemically similar. For example, if we add ``José'' as a context phrase, the ASR system might falsely recognize ``O say can you see'' as ``José can you see''. This issue is particularly acute for a general ASR system that is not restricted to a particular domain. As a result, this drawback makes biased models less competitive, as the benefits gained may be outweighed by the negative impact on overall performance.



In this paper, we focus on investigating how to mitigate the performance degradation seen in common scenarios caused by contextually biased Conformer-Transducer models.
Previous research~\cite{filter} has indicated that filtering out irrelevant context phrases can reduce the negative impact of biasing. However, previous filtering methods are based on phonemic similarity, thus can sometimes miss acoustically similar words, and compromise the model's ability to prevent false alarms. Furthermore, the two-stage filtering approach proposed in~\cite{filter} relies on the results of the first pass to filter context phrase candidates, which can hinder the model's ability to perform streaming recognition. Additionally, the repetition of joint operations for Transducer models in different stages can complicate the inference pipeline and increase the computation costs.
To address the filtering accuracy issue, we hypothesize that to differentiate those acoustically similar words, acoustic information is not enough, and semantic information is required as well. 
We introduce an Entity Detector (ED) that predicts the occurrence of context words with the multi-head attention matrix of the biased encoder and predictor output. To maintain a streaming transducer pipeline and avoid the additional joint cost, we use a single-staged approach -- the filtering of the context list does not require the posterior matrix from previous stages and the biased encoder and predictor output we used instead can be calculated in a streaming way.

The output of the introduced ED module is used as a switch for the list of context phrases in the inference process so that the context embedding list will only be used if context phrases appear in the speech during the inference. We also investigate two different strategies of the proposed ED module. The first one is called the Predictor based Entity Detector (P-ED), which uses only the predictor output and has the advantage of less inference complexity. The other one is called the Encoder-Predictor based Entity Detector (EP-ED), which uses the encoder bias output and the predictor bias output to perform multi-headed attention. Because it uses biased context information as input, biasing operations are needed even for non-bias frames, thus leading to a slight increase in inference latency.

Experiments on Librispeech~\cite{librispeech} and our internal voice assistant dataset show that our approach can achieve up to 6.7\% and 20.7\% relative reduction in word error rate (WER) compared to the baseline respectively, mitigating up to 96.7\% and 84.9\% of the relative WER increase for common cases. Furthermore, our approach has a minimal performance impact in personalized scenarios while maintaining a streaming inference pipeline with negligible real-time factor (RTF) increase.


\begin{table*}[h]
\centering
\setlength{\belowcaptionskip}{0.05cm} 
\setlength{\abovecaptionskip}{0.15cm} 
\caption{Performance (WER\%) of different biasing list size N on the LibriSpeech benchmark.}
\label{librispeech}
\resizebox{1.0\textwidth}{!}{
\begin{tabular}{@{}lccccccccc@{}}
\cline{1-10}
\multicolumn{1}{l|}{\multirow{2}{*}{Model}} & \multicolumn{1}{c|}{\multirow{2}{*}{Test Set}}
  & \multicolumn{2}{c|}{N=0} &
  \multicolumn{2}{c|}{N=20} &
  \multicolumn{2}{c|}{N=50} &
  \multicolumn{2}{c}{N=100} \\ 
  \multicolumn{1}{l|}{} &
  \multicolumn{1}{c|}{} &
  test-clean &
  \multicolumn{1}{c|}{test-other} &
  test-clean &
  \multicolumn{1}{c|}{test-other} &
  test-clean &
  \multicolumn{1}{c|}{test-other} &
  test-clean &
  test-other \\ \cline{1-10}
\multicolumn{1}{l|}{C-T} &\multicolumn{1}{c|}{Personalized/Common} &
  4.39 &
  \multicolumn{1}{c|}{9.05} &
  4.39 &
  \multicolumn{1}{c|}{9.05} &
  4.39 &
  \multicolumn{1}{c|}{9.05} &
  4.39 &
  9.05 \\ \cline{1-10}
  \multicolumn{1}{l|}{CATT} &
\multicolumn{1}{c|}{\multirow{3}{*}{Personalized}} &
  4.40  &
  \multicolumn{1}{c|}{9.07} &
  3.92 &
  \multicolumn{1}{c|}{8.42} &
  3.94 &
  \multicolumn{1}{c|}{8.53} &
  4.04 &
  8.62 \\
  \multicolumn{1}{l|}{~+P-ED} &
\multicolumn{1}{c|}{} &
  4.37 &
  \multicolumn{1}{c|}{8.89} &
  4.04 &
  \multicolumn{1}{c|}{7.90} &
  3.97 &
  \multicolumn{1}{c|}{8.10} &
  \textbf{3.87} &
  8.15 \\
  \multicolumn{1}{l|}{~+EP-ED} &
\multicolumn{1}{c|}{} &
  \textbf{4.35} &
  \multicolumn{1}{c|}{\textbf{8.83 }} &
  \textbf{3.91} &
  \multicolumn{1}{c|}{\textbf{7.89}} &
  \textbf{3.94} &
  \multicolumn{1}{c|}{\textbf{8.06}} &
  3.90 &
  \textbf{8.13} \\ \cline{1-10}
  \multicolumn{1}{l|}{CATT} &
\multicolumn{1}{c|}{\multirow{3}{*}{Common}} &
  
  4.40 &
  \multicolumn{1}{c|}{9.07} &
  4.65 &
  \multicolumn{1}{c|}{9.37} &
  4.81 &
  \multicolumn{1}{c|}{9.41} &
  4.85 &
  9.56 \\
  \multicolumn{1}{l|}{~+P-ED} &
\multicolumn{1}{c|}{} &
  4.37 &
  \multicolumn{1}{c|}{8.89} &
  4.41 &
  \multicolumn{1}{c|}{9.02} &
  4.45 &
  \multicolumn{1}{c|}{9.12} &
  4.72 &  
  9.42 \\
  \multicolumn{1}{l|}{~+EP-ED} &
\multicolumn{1}{c|}{} &
  \textbf{4.35} &
  \multicolumn{1}{c|}{\textbf{8.83}} &
 \textbf{4.40}&
  \multicolumn{1}{c|}{\textbf{8.99}} &
  \textbf{4.43} &
  \multicolumn{1}{c|}{\textbf{9.09}} &
  \textbf{4.71} &
  \textbf{9.41} \\ \cline{1-10}
\end{tabular}}
\vspace{0cm} 
\end{table*}
\section{Method }

In this section, we introduce the overall framework of our base ASR system, as shown in Figure 1 (a) and (b). Generally, we add an additional Entity Detector module to the base CATT model.

\subsection{Predictor Entity Detector (P-ED)}
Because context words may cause false alarms, if we know that the current utterance contains no context phrases from the entity list, then it is better to inference with no entity list at all (an empty entity list). Therefore, we aim to design a module that can predict the occurrence of the context words, and use the prediction to assist biased inference.

We use a Context Encoder to convert a list of context words to context embedding.  Each contextual word or phrase $w_k$ is first represented as a sub-word for English and a character for Mandarin, and then fed into the context encoder to produce fixed dimensional vector representations $h_k^{CE}$.
Then, the Encoder and Predictor Biasing Layers (Figure 1(e)) are introduced to allow the model to learn the relationship between contextual phrases and speech. This relationship is learned with a multi-head attention (MHA) layer. The audio embedding $h_t^E$ serves as the query vector, while the contextual phrase embedding $h_k^{CE}$ refers to the keys and values vectors. 
The Combine module consists of a LayerNorm layer, a concatenation operation, and a feed-forward projection layer. This process can be described as:
\begin{equation}
h^{concat}_u=[\text{LayerNorm}(h^{PB}_u),\text{LayerNorm}(h_u^{P})]~,
\end{equation}
\begin{equation}
h^{CAP}_u=\mathrm{FeedForward}(h^{concat}_t)
\end{equation}
To predict the occurrence of the context words, we add a Predictor based Entity Detector (P-ED) as shown in Figure (a) and (c). Inspired by the biasing layers, P-ED is an MHA layer designed to predict the occurrence of context phrases according to Predictor output $h_u^{CAP}$. Specifically, the predictor output $h_u^{CAP}$ is used as the query vector, and the contextual phrase embedding $h_k^{CE}$ is used as the key and value. More specifically, the query, key, and value are calculated as:
\begin{equation}
\begin{aligned}
Q^{ed} &= \sigma\big(XW^{ed,q} + 1(b^{ed,q}_{i})^\top\big), \\
K^{ed} &= \sigma\big(CW^{ed,k} + 1(b^{ed,k}_{i})^\top\big),\\
V^{ed} &= \sigma\big(CW^{ed,v} + 1(b^{ed,v}_{i})^\top\big), 
\end{aligned}
\end{equation}
where $X=[h_1^{CAP},...,h_U^{CAP}]^\top$ is the concatenated predictor bias output of an utterance, and $U$ is the number of tokens. $C=[h_1^{CE},...,h_k^{CE}]^\top$ is the  context embeddings, where $K$ is the number of context phrases. The entity detection embedding  $h_t^{ed}$ is calculated via multihead attention:

\begin{equation}
H^{ed} = \text{Softmax}\bigg(\frac{Q^{ed}(K^{ed})^\top}{\sqrt{d}}\bigg)V^{ed},
\end{equation}
where the scaling factor $\sqrt{d}$ is for numerical stability. 
The entity detection embedding  $h_t^{ed}$ is then used to perform a 2-class classification task against the ground truth label of the text via a linear layer. A cross-entropy loss, $L_{bias}$  is calculated. This process can be described as:

\begin{equation}
h^{predict}_t=\text{FeedForward}(h^{ed}_t)~.
\end{equation}
The final joint loss function for our model is:
\begin{equation}
L= L_{transducer}+\lambda_1 L_{bias},
\end{equation}
where $\lambda_1$ is a weight hyperparameter we set as $0.4$ in all  our experiments.

\subsection{Encoder-Predictor based Entity Detector (EP-ED)}
Because the predictor embedding may lack the relevant acoustic information for the ED to deduce the presence of a context phrase, we design another framework to incorporate acoustic information as well. 
Different from Section 2.1, the EP-ED do not take context embedding ${h}_{t}^{CE}$ and predictor embedding ${h}_{t}^{PE}$ as input, instead, it uses biased predictor embedding ${h}_{t}^{PB}$ as query and biased  encoder embedding ${h}_{t}^{EB}$ as key and value.
Figure 1 (b) and (d) illustrates how we use both audio embeddings and label embeddings to detect the occurrence of context word. Specifically, use biasing layers, one taking the biased encoder embeddings ${h}_{t}^{EB}$as keys and values, and the biased predictor embeddings ${h}_{u}^{PB}$ as queries to obtain the entity prediction.

Similar to that in Section 2.1, to predict the context phrase, the context-aware predictor-encoder embedding is then used for a 2-class classification task. This enables the ED to predict whether the current token is contained within a context word or not. 

\subsection{Adaptive Contextual Inference }

For the baseline CATT method, every frame of the encoder out need to be biased using encoder bias during inference, and with every none blank token, a predictor out $h_t^{EP}$ is generated and used to compute predictor biased output $h_t^{PB}$. With the introduction of the entity detector, if it predicts that a context word is present,  then the context encoder will use the context embedding of  $w_i$, otherwise, it will use the context embedding of an empty list.

For the P-ED approach, biased encoder embedding $h_t^{EB}$ is not required when the Entity Detector predicts that context phrases are not present in the current token. Thus only the computation is needed only for tokens where  an entity is detected.

For the EP-ED approach, the inference process is a bit more complex. Because it integrates the context-aware acoustic information from the encoder, it comes Because EP-ED requires both biased encoder embedding and biased predictor embedding. For every non-blank token, a prediction is made. If no context phrase is detected, the predictor bias and encoder bias are recomputed using the context embedding of an  empty context word list.

\begin{table*}[h]
\centering
\setlength{\belowcaptionskip}{0.05cm} 
\setlength{\abovecaptionskip}{0.15cm} 
\caption{Ablation study on Librispeech benchmark}
\label{ablation}
\resizebox{0.85\textwidth}{!}{
\begin{tabular}{l|cccccccc}
\hline

\multicolumn{1}{c|}{}                       & \multicolumn{2}{c|}{N=20}                                 & \multicolumn{2}{c|}{N=50}                                  & \multicolumn{2}{c|}{N=20}                           & \multicolumn{2}{c}{N=50}                    \\ 
\multicolumn{1}{c|}{}                       & test-clean           & \multicolumn{1}{c|}{test-other}    & test-clean           & \multicolumn{1}{c|}{test-other}     & test-clean     & \multicolumn{1}{c|}{test-other}    & test-clean           & test-other           \\ \hline
CATT                                        & 3.92                 & \multicolumn{1}{c|}{8.42}          & 3.94                 & \multicolumn{1}{c|}{8.53}           & 4.65           & \multicolumn{1}{c|}{9.37}          & 4.81                 & 9.41                 \\ \hline
~~+ P-ED                                      & 4.04                 & \multicolumn{1}{c|}{7.90}          & 3.97                 & \multicolumn{1}{c|}{8.10 }  & \textbf{4.41}  & \multicolumn{1}{c|}{\textbf{9.02}} & \textbf{4.45}        & \textbf{9.12}        \\
- detector                                  & \textbf{3.77}        & \multicolumn{1}{c|}{\textbf{7.83}} & \textbf{3.87}        & \multicolumn{1}{c|}{\textbf{7.99}}           & 4.62           & \multicolumn{1}{c|}{9.32}          & 4.73                 & 9.38                 \\
- 50\% detector                             & 4.12                 & \multicolumn{1}{c|}{8.31}          & 4.19                 & \multicolumn{1}{c|}{8.40}           & 4.53           & \multicolumn{1}{c|}{9.15}          & 4.57                 & 9.19                 \\ \hline
+ EP-ED                                     & 3.91                 & \multicolumn{1}{c|}{7.89}          & 3.93                 & \multicolumn{1}{c|}{8.06}           & \textbf{4.37}  & \multicolumn{1}{c|}{\textbf{8.99}} & \textbf{4.43}        & \textbf{9.09}        \\
- detector                                  &  \textbf{3.79}                 & \multicolumn{1}{c|}{ \textbf{7.87}}          &  \textbf{3.89}                 & \multicolumn{1}{c|}{\textbf{7.96}}  & 4.59           & \multicolumn{1}{c|}{9.30}          & 4.69                 & 9.34                 \\
- 50\% detector                             & 4.13                 & \multicolumn{1}{c|}{8.29}          & 4.21                 & \multicolumn{1}{c|}{8.39}           & 4.52           & \multicolumn{1}{c|}{9.14}          & 4.57                 & 9.17                 \\ \hline
\end{tabular}
}
\vspace{0cm} 
\end{table*}

\section{Experiments}

\subsection{Data}

\textbf{LibriSpeech:} We used the LibriSpeech corpus~\cite{librispeech} for some of our experiments. It contains 960 hours of English audiobook recordings. The training was performed
using the full training set of 960h, with SpecAugment~\cite{specaugment} applied. 
We evaluate the models on the dev and test sets provided with the LibriSpeech corpus: \textit{dev-clean}, \textit{dev-other }\textit{, test-clean} and  \textit{test-other}.
For Librispeech,  \textit{test-clean} and  \textit{test-other} are used to create both \textit{common} and \textit{personalized} test sets. The only difference between  \textit{common} and \textit{personalized} is the entity list.
For the \textit{personalized} set, entity lists for the LibriSpeech corpus from~\cite{2021nnbias3} were used. The entity lists are created by selecting  words  from each utterance that appear in a rare-word list. and then adding a certain number of distractor words.
For \textit{common} set, the entity list contains only distractor words.

\textbf{Voice assistant dataset:} To validate the model’s performance, a larger in-house dataset containing 15,000 hours anonymized Mandarin speech collected from a speech assistant product was used. The \textit{personalized} and\textit{ common} sets are created individually. The \textit{personalized} test set contains approximately 500 utterances with 50 contextual words. Most of the contextual words are personal names. 
For the \textit{common} set, there are about 10000 utterances, the same entity list for \textit{common} set was used, and the utterances contain no context phrases from the provided  entity list.

\subsection{Experimental Setups}

Our proposed method and the baseline Conformer Transducer model (referred to as C-T) shares the following parameters:
 The encoder is a 12-layer conformer. Each Conformer block consists of 4-head 256-dim multi-head attention block, The predictor is a  2-layer LSTM. Both the encoder and predictors have an embedding size of 256. The joint network is a fully-connected feed-forward component with one hidden layer followed by a Tanh activation function. 

For models other than the the C-T baseline, we used the LSTM context encoder as mentioned in Section 2.1. It consists of a 1-layer BLSTM with a dimension of 256. The bias layer is a transformer block with an embedding size of 256 and 4 attention heads. The linear layer in the entity detector keeps a 256-dimensional input and output.


In addition to the word error rate (WER) for the English corpus and character error rate(CER) for the mandarin corpus, results were further evaluated via the biased label character error rate (L-CER) which is defined by:

\begin{equation}
\text{L-CER}=\frac{\text{EditDistance}(ref, hyp)}{\max(\text{len}(ref), \text{len}(hyp))}
\end{equation}
where ref is the ground truth of the context phrase label, hyp is the prediction of ED. L-CER is a direct assessment of the ED prediction, compared with WER metrics, we use it as a more direct indicator of  the performance of the ED module.

\subsection{Contextual ASR Accuracy}


\begin{table}[!t]
\centering
\setlength{\belowcaptionskip}{0.05cm} 
\setlength{\abovecaptionskip}{0.15cm} 
\caption{Performance(L-CER\%) on Librispeech, TO stands for test other, TC stands for test clean, Pers. stands for Personalized, Comm. stands for Common.}
\label{fusion}
\resizebox{0.47\textwidth}{!}{

\begin{tabular}{l|c|cc|cc|cc}
\hline
\multicolumn{1}{c|}{\multirow{2}{*}{Model}} & \multicolumn{1}{c|}{\multirow{2}{*}{Test Set}} & \multicolumn{2}{c|}{N=20}     & \multicolumn{2}{c|}{N=50}       & \multicolumn{2}{c}{N=100}       \\ \cline{3-8} 
\multicolumn{1}{c|}{}                       & \multicolumn{1}{c|}{}                          & TC            & TO            & TC             & TO             & TC             & TO             \\ \hline
\multirow{2}{*}{CATT + P-ED}                &   \multirow{2}{*}{Pers.}                                   & 10.98         & 9.72          & 19.34          & 18.36          & 43.24          & 35.42          \\
                                            &                                                & \textbf{7.04} & \textbf{8.53} & \textbf{15.84} & \textbf{13.57} & 39.84          & \textbf{37.36} \\ \hline
\multirow{2}{*}{CATT + P-ED}                & \multirow{2}{*}{Comm.}                                       & 4.97          & 3.40          & 17.82          & 13.86          & 91.52          & 85.87          \\
                                            &                                                & \textbf{4.38} & \textbf{2.02} & \textbf{13.62} & \textbf{10.72} & \textbf{90.93} & \textbf{83.25} \\ \hline
\end{tabular}

}
\vspace{-0.3cm} 
\end{table}

In this section, we tested both entity detection methods and non-contextual baselines for both \textit{personalized} and  \textit{common} sets. As shown in Table \ref{librispeech}, 
Experiments for the \textit{common} set show a significant decrease in WER for the common dataset, with up to 6.7 \% relative WERR for the test other with an entity list of 20. Significantly reducing the WER increase incurred by bias word by 96.7\%, as the entity detector has a low false alarm rate, and can predict the occurrence of context phrases with an L-CER under 5\% as shown in Table 3.

The result of the \textit{personalized} set showed a WER reduction for all entity list sizes compared with CATT baseline. The difference is especially significant in the test-other set. This is unlikely caused by the additional parameters, since the increase in model size is negligible.   We hypothesize that this is caused by the introduced loss function helping to supervise the gradient descent process and thus leading to better performance. This hypothesis is further validated in the ablation study. 


However, when the size of the biasing list grows over 100, both P-ED and EP-ED showed a significant increase in L-CER, because when the biasing list grows too large, the ED tends to predict always on, thus reducing the performance for common scenarios. This suggests that the model is sensitive to the size of the biasing list, which is related to the number of context words that are used during training. 

\subsection{Ablation Study}

As mentioned in Section 2.1, ED is introduced to predict the occurrence of context phrases. To verify it, we removed features from removing features from P-ED and EP-ED (- detection means always attending to context phrase regardless of TD prediction, -50\% detection means attending to context phrase with 50\% chance, regardless of TD prediction). Table \ref{ablation} presents the result.
 Both P-ED and EP-ED show improvement over the 50\% bias, with 
less WER in both personalized and common scenarios. This implies that through training both entity detectors have learned the ability to predict context words. 
When compared to a always-on bias (-detector), it shows that even without using the entity detector during inference, both entity detector methods show a significant improvement over the base CATT method, probably due to the introduced supervision helping the model to be better optimized during training via the introduced supervision. Both P-ED and EP-ED, improve  \textit{common} set WERs with EP-ED have better performance over P-ED overall, compared with always-on bias.

\subsection{RTF Analysis}
To evaluate the performance of our proposed filter module, we compared the runtime RTF of the ASR system with and without the biasing module. We conducted the experiment on a 3.40GHz AMD Ryzen 5950X CPU using a single thread. As shown in Table 4, the results indicate that the addition of the proposed module leads to a minor increase in RTF when compared to the baseline system. This is mainly due to the fact that the encoder bias $h_t^{EB}$ is unnecessary when the proposed module predicts the absence of a contextual phrase. Additionally, the computational cost of the prediction process is negligible. However, when comparing the RTF of the EP-ED system with the baseline, we observed a slight increase in RTF. This is because the EP-ED system requires the computation of every context-aware acoustic embedding $h_t^{EB}$ and $h_t^{PB}$, as in CATT, to make a prediction.

\begin{table}[!t]
\centering
\setlength{\belowcaptionskip}{0.05cm} 
\setlength{\abovecaptionskip}{0.15cm} 
\caption{Performance of Entity Detector on in-house dataset}
\label{gigaspeech}
\resizebox{0.47\textwidth}{!}{

\begin{tabular}{l|cc|c|c}
\hline
\multicolumn{1}{c|}{\multirow{2}{*}{Model}} & \multicolumn{2}{c|}{CER(Personalized)} &  CER(Common)    & \multirow{2}{*}{RTF} \\
\multicolumn{1}{c|}{}                       & N=0           & N=50                  & N=50          &                      \\ \hline
C-T                                         & 15.75         & 15.75                 & \textbf{6.64}          & 0.108                \\
CATT                                        & 15.89         & \textbf{6.93}                  & 8.82          & 0.128                \\
~+P-ED                                 & 16.03         & 7.33                  & 7.19          & 0.130                \\
~+EP-ED                                & 14.43         &  7.02         & 6.97 & 0.135                \\ \hline
\end{tabular}
}

\vspace{-0.3cm} 
\end{table}

\subsection{Test on Voice Assistant Dataset}

We further validate our result on an in-house voice assistance dataset. 
For the common set, When utilizing a biased list, the P-ED achieved a 16.4\% CERR  and the EP-ED achieved a 20.7\% CERR compared to the baseline CATT. Mitigating the CER increase introduced by biasing modules by 84.9\% and 74.7\% respectively.  When not using a bias list, the performance has no significant difference compared to the baseline.
Both P-ED and EP-ED did not incur significant performance degradation, having 5.4\% and 1.2\% relative CER increase respectively.

\section{Conclusions}

In this work, we studied how to reduce the WER increase introduced by biasing modules in common scenarios. We proposed a novel adaptive method based on entity detector. The entity detector predicts context phrases contained in the utterances and calculates cross-entropy loss on the prediction results to train the ED to filler out the irrelevant entity list during inference.  Compared with previous methods, our approach has better performance overall, achieving significant word error rate reduction for common scenarios.


\nocite{*}
\bibliography{mybib.bib} 
\bibliographystyle{IEEEtran} 
\end{document}